\documentclass{article}
\usepackage{times}
\usepackage{amsmath}
\usepackage{amssymb}
\usepackage{array}
\usepackage{epsfig}
\usepackage{pstricks}

\renewcommand{\theequation}{\thesection.\arabic{equation}}
\newcommand{\be}{\begin{equation}}
\newcommand{\ee}{\end{equation}}

\title{ Dyons and Magnetic Monopoles  Revisited } 
\author{
Edward A. Olszewski  \\ 
 {\em Department of Physics} \\ 
 {\em University of North Carolina at Wilmington} \\  
 {\em Wilmington, North Carolina 28403-5606} \\
{\em email: olszewski@uncw.edu}
}
\date{} 

\begin{document}

\bibliographystyle{plain}

\maketitle

\begin{abstract}
We construct dyon solutions  in SU($N$)  with topological electric and magnetic charge.  Assuming a $|\boldsymbol{\Phi}|^4$--\:like potential for the Higgs field we show that the mass of the dyons is relatively insensitive to the coupling parameter $\lambda$ characterizing the potential.  We then apply the methodology of constructing dyon solutions in SU($N$)  to G2.   In order to define the electromagnetic field consistently in the manner that we propose we find that dyon solutions exist only when G2 is considered under the action of its maximal and regular subgroup SU(3).  In this case we find two different types of dyons, one of which  has properties identical to dyons  in SU(3).  The other dyon has some properties which are seemingly atypical, e.g.\@ the magnetic charge $g_m = 4 \pi\: 3/e$, which differs from the 't Hooft/Polyakov monopole where $g_m = 4 \pi\: 1/e$.  
\end{abstract}

\setcounter{page}{1}
\setcounter{section}{0}
\setcounter{equation}{0}
\section{Introduction}
\label{sect1}
The subject of magnetic monopoles has intrigued and fascinated the physics community dating back to the early twentieth century.  Dirac, first, piqued interest in this subject by providing a theoretical argument demonstrating that  the existence of  magnetic monopoles requires not only that electric charge be quantized but also that the electric and magnetic couplings be inversely proportional to each other, i.e. weak/strong duality.  Subsequently, 't Hooft \cite{thooftg76} and Polyakov showed that  within the context of the spontaneously broken Yang-Mills gauge theory SO(3) magnetic monopole solutions of finite mass must necessarily exist and furthermore possess an internal structure.  Consequently, Montonen and Olive conjectured that there was an exact weak/strong, electromagnetic duality for the spontaneously broken SO(3) gauge theory \cite{montonenc77}. More recently, this conjecture has become credible in the broader context of $N=2$ or $N=4$ Super Yang-Mills theories~\cite{harveyj96}.

Our purpose here is to investigate magnetic monopole--like solutions, or more specifically dyon solutions which possess both topological electric and magnetic charge.  We introduce a methodology of constructing such solutions within the context of an arbitrary gauge group, emphasizing, in particular, the  groups SU(N) and G2.   In Section~\ref{sect2_1} we introduce a necessary condition which we consider a prerequisite for categorizing dyon solutions.   Based on the condition we  construct dyon solutions for the gauge group SU($N$).   Finally, we apply the construction, specifically, to SU(3) and  G2 emphasizing differences and commonalities between these  two types of solutions and those of  SU(2).

Concerning conventions we adopt those of Harvey~\cite{harveyj96} with the exception that the Levi-Civit\`{a} symbol $\varepsilon_{0123} = \varepsilon_{123} = 1$.  We summarize the other relevant conventions:  the Minkowski signature is $(+---)$; Greek letters denote space time indices, i.e. 0, 1, 2, 3, while Roman letters denote either the spatial indices  1, 2, 3 or the indices of  the generators of the  gauge group.  Also the gauge coupling is denoted $e$.  We  employ Lorentz-Heaviside units of electromagnetism  so that $c=\hbar=\epsilon_0=\mu_0=1$. One  implication is that the Dirac quantization condition is $e\:g_m = (4 \pi) n/2$, $g_m$ being the magnetic charge and $n$ being an integer.
\section{The Electromagnetic Field}
\label{sect2_1}

In this section we adopt  the definition of the electromagnetic field first introduced by 't Hooft~\cite{thooftg76} in the context of the gauge group SO(3).  Furthermore, we show that this definition of the electromagnetic can be consistently applied to an arbitrary gauge field for a particular gauge group when a specific condition, which we derive, is satisfied.

%We then show that this definition of the electromagnetic field can be applied to other 
%gauge groups when  a condition which we derive in this section is satisfied. 

Consider the Yang--Mills--Higgs Lagrangian:
\be
\mathcal{L} = -\frac{1}{4} \boldsymbol{F}_{\mu \nu} \cdot \boldsymbol{F}^{\mu \nu} + \frac{1}{2} D_\mu \boldsymbol{\Phi} \cdot D^\mu \boldsymbol{\Phi} - V(\boldsymbol{\Phi}\cdot \boldsymbol{\Phi})      \: ,\label{eq2_1}
\ee
where 
\be
\boldsymbol{F}_{\mu \nu} = \partial_\mu \boldsymbol{A}_{\nu} - \partial_\nu \boldsymbol{A}_{\mu} - i e\:  \boldsymbol{A}_{\mu} \wedge \boldsymbol{A}_{\nu} \:. \label{eq2_1b} 
\ee
The Higgs field $\boldsymbol{\Phi}$ is an isoscalar transforming according to the adjoint representation of the gauge group so that its covariant deriviative is
\be
D_\mu \boldsymbol{\Phi} = \partial_\mu \boldsymbol{\Phi}  - i e\: \boldsymbol{A}_{\mu} \wedge \boldsymbol{\Phi}  \:.  \label{eq2_1c}
\ee  
The generators of the algebra, $T_a$ have been chosen so that Tr$(T_a T_b) = (1/2) \delta_{a b}$. For two fields, $\boldsymbol{A}$ and $\boldsymbol{B}$,  transforming as the adjoint representation of the gauge group  $\boldsymbol{A }\cdot \boldsymbol{B}$ is defined as
\be
\boldsymbol{A} \cdot \boldsymbol{B} \equiv 2\: \text{Tr} (\boldsymbol{A}\boldsymbol{B}) =  \: A^a B^a  \:. \label{eq2_2}
\ee
One possible definition for the electromagnetic field tensor $F$ is the gauge invariant quantity $F_{\mu \nu} = \boldsymbol{F}_{\mu \nu} \cdot   \boldsymbol{\Phi} $.  This does not suffice because, in general,   $F_{\mu \nu}$   neither satisfies the Maxwell equations  nor the Bianchi identity $dF=0$.  When considering the gauge group SO(3) 't Hooft proposed the following modification, 
\be F_{\mu \nu} =  \boldsymbol{F}_{\mu \nu} \cdot   \boldsymbol{\bar{\Phi}} - \frac{1}{i e}\:   D_\mu \boldsymbol{\bar{\Phi}} \wedge  D_\nu \boldsymbol{\bar{\Phi}}  \cdot \boldsymbol{\bar{\Phi}}  \: , \label{eq2_3}
\ee
where  $\boldsymbol{\bar{\Phi}} \cdot \boldsymbol{\bar{\Phi}} = C^2$.  Here $C$ is a constant which 't Hooft sets equal to one.  We now show  in order that $dF_{\mu \nu}=0$ the following condition is sufficient,
\be
D_\mu \boldsymbol{\bar{\Phi}}  =  \boldsymbol{\bar{\Phi}}  \wedge   (\boldsymbol{\bar{\Phi}} \wedge  D_\mu \boldsymbol{\bar{\Phi}}) \:.  \label{eq2_4} 
\ee   
First, note that this condition is satisfied for SO(3) or SU(2) as can be seen, straightforwardly, from the following heuristic arguement.  The algebra of SO(3) and SU(2) is essentially the same as the vector  cross product in three dimensional space.  Thus, if we consider  $\boldsymbol{\bar{\Phi}}$  to be the element of the SU(2) algebra, $T_z$,  then we can assume  that $D_\mu \boldsymbol{\bar{\Phi}}$, which is ``perpendicular'' to $T_z$,  is   $D_\mu \boldsymbol{\bar{\Phi}} = -i e A T_x$.  Consequently, the right-hand side of Eq.~\ref{eq2_4} is replaced by $T_z  \wedge (T_z \wedge (-i e) A T_x)  =  -i e A T_x = D_\mu \boldsymbol{\bar{\Phi}}$. Here we have used the fact that the SU(2) algebra satisfies  $T_i \wedge T_j = i \epsilon_{ijk} T_k$.

We now derive the condition~\ref{eq2_4}.  The Bianchi identity can be expressed as
\be
 dF = \partial_\gamma F_{\mu \nu} dx^\gamma \wedge dx^\mu \wedge dx^\nu = D_\gamma F_{\mu \nu} dx^\gamma \wedge dx^\mu \wedge dx^\nu \: \label{eq2_5}
\ee
Since $ \boldsymbol{F}_{\mu \nu} $ is a two-form, Eq.~\ref{eq2_5} can be re-expressed as
\be
\begin{split}
dF =  \{ D_\gamma \boldsymbol{F}_{\mu \nu} & \cdot \boldsymbol{\bar{\Phi}} +  
    \boldsymbol{F}_{\mu \nu} \cdot D_\gamma \boldsymbol{\bar{\Phi}} 
- \frac{1}{i e} D_\gamma  \boldsymbol{\bar{\Phi}} \cdot D_\mu \boldsymbol{\bar{\Phi}} \wedge  D_\nu \boldsymbol{\bar{\Phi}} \\
& - \frac{1}{i e} (D_\gamma \wedge D_\mu )\boldsymbol{\bar{\Phi}} \wedge  D_\nu \boldsymbol{\bar{\Phi}}  \cdot \boldsymbol{\bar{\Phi}}  \} \: dx^\gamma \wedge dx^\mu \wedge dx^\nu   \:.  \label{eq2_6}
\end{split}
\ee
Using the Jacobi identity, the cyclic property of the trace, and $(D_\gamma \wedge D_\mu) \boldsymbol{\bar{\Phi}} = - i e\: \boldsymbol{F}_{\gamma \mu} \wedge \boldsymbol{\bar{\Phi}}$, we can re-express Eq.~\ref{eq2_6} as
\be
\begin{split}
dF =\{D_\gamma \boldsymbol{F}_{\mu \nu} & \cdot \boldsymbol{\bar{\Phi}} +  
    \boldsymbol{F}_{\mu \nu} \cdot D_\gamma \boldsymbol{\bar{\Phi}} 
- \frac{1}{i e} D_\gamma  \boldsymbol{\bar{\Phi}} \cdot D_\mu \boldsymbol{\bar{\Phi}} \wedge  D_\nu \boldsymbol{\bar{\Phi}} \\
& - \boldsymbol{F}_{\mu \nu} \cdot \boldsymbol{\bar{\Phi}}  \wedge   (\boldsymbol{\bar{\Phi}} \wedge  D_\gamma \boldsymbol{\bar{\Phi}}) \}\: dx^\gamma \wedge dx^\mu \wedge dx^\nu   \:.  \label{eq2_7}
\end{split}
\ee
The first term vanishes because of the Jacobi identity.  The second and fourth terms cancel because  the condition given in Eq.~\ref{eq2_4} is satisfied. The third term vanishes for the following reason. Without loss of generality we assume that $\boldsymbol{\bar{\Phi}}$ is an element of the Cartan subalgebra  multiplied by a constant. Furthermore, Eq.~\ref{eq2_4} is equivalent to the following condition imposed on the root vectors $E_{\boldsymbol{\alpha}}$,    i.e.\@ the raising and lowering operators of the weights,\footnote{The equivalence of the two conditions can be shown by assuming a general form for  $\boldsymbol{A}_{\mu}$, i.e.\@     $\boldsymbol{A}_{\mu} = A^a_\mu\: t_a$  where $t_a$ is an $E_{\boldsymbol{\alpha}}$ or an element of the Cartan subagebra, and performing the calculations in Eq.~\ref{eq2_4} directly.}  
\be
\boldsymbol{\bar{\Phi}}  \wedge  E_{\boldsymbol{\alpha}} =  \pm E_{\boldsymbol{\alpha}} \: \text{or}\: 0 \:. \label{eq2_8}
\ee
Consequently, if Eq.~\ref{eq2_8} is satisfied, $ D_\mu \boldsymbol{\bar{\Phi}} \wedge  D_\nu \boldsymbol{\bar{\Phi}} \propto \boldsymbol{\bar{\Phi}}$ or commutes with  $\boldsymbol{\bar{\Phi}}$; otherwise, there would exist a root vector which does not satisfy  Eq.~\ref{eq2_8}. Since $D_\gamma  \boldsymbol{\bar{\Phi}}$ does not commute with $\boldsymbol{\bar{\Phi}}$ (othewise, it would be zero) and is ``perpendicular'' to $\boldsymbol{\bar{\Phi}} $, the third term vanishes. Thus, we have shown that Eq.~\ref{eq2_4} is sufficient to imply  $dF = 0$.

\setcounter{equation}{0}
\section{Dyon Solutions in SU($N$)}
\label{sect5}

We now construct dyon solutions for SU($N$). The choice of  $\boldsymbol{\bar{\Phi}}$ must be in accordance with   Eq.~\ref{eq2_8}.  To this end we consider the  root vector $E_{\boldsymbol{\alpha}}$ of SU($N$) whose  root is
\be
\boldsymbol{\alpha} =  (0, 0, \cdots, \alpha_2, \alpha_1) \:, \label{eqa_1} 
\ee
 where 
\be
\begin{split}
\alpha_{1} &=  \sqrt{\frac{N}{2(N-1)}}   \\
\alpha_{2} &= \sqrt{\frac{N-2}{2(N-1)}}       \:. \label{eqa_2}
\end{split}
\ee
Let $H_{N-i}$ ($i=1, 2$) be those elements of the Cartan subalgebra for which
\be
H_{N-i} \wedge E_{\boldsymbol{\alpha}} = \alpha_i E_{\boldsymbol{\alpha}} \:.   \label{eqa_3}
\ee
We choose 
\be
\boldsymbol{\bar{\Phi}}  = \frac{1}{\alpha_1} H_{N-1} \: \label{eqa_2a}
\ee
This choice of $\boldsymbol{\bar{\Phi}}$ satisfies the condition \ref{eq2_8} for all root vectors of SU($N$).  For example, in the case of SU($3$) this corresponds to the choice  $\boldsymbol{\bar{\Phi}} = \frac{1}{\alpha_1}H_2$ as can be seen in Figure~\ref{fig1}. 

As in the work of Harvey~\cite{harveyj96} we assume that the potential  $V(\boldsymbol{\Phi}\cdot \boldsymbol{\Phi})$ to be of a form such that the vacuum expectation of $\Phi$ is non-zero.  When a specific form of  $V(\boldsymbol{\Phi}\cdot \boldsymbol{\Phi})$ is required we use 
\be
V(\boldsymbol{\Phi}\cdot \boldsymbol{\Phi}) = \frac{\lambda}{8} ( \boldsymbol{\Phi}\cdot \boldsymbol{\Phi} -v^2)^2  \: .  \label{eq2_10}
\ee      
In general, the Higgs vacuum is defined to be the set of all $\boldsymbol{\Phi}$ such that $V(\boldsymbol{\Phi}\cdot \boldsymbol{\Phi}) = 0$. For the  specific form of the potential given by Equation~\ref{eq2_10} this implies that $\boldsymbol{\Phi}\cdot \boldsymbol{\Phi} = v^2$ and consequently that the various vacuum states can be labeled by $v$.

Define\footnote{Although $|\boldsymbol{\alpha}| = 1$ for SU($N$), we, nonetheless, explicitly include $|\boldsymbol{\alpha}| $ in subsequent formulae in anticpation of generalizing  these results to G2.} 
\be 
\begin{split}
T_\pm  = & \frac{E_{\pm \alpha}}{|\boldsymbol{\alpha}|} \\  
T_z  = & \frac{\alpha_2 H_{N-2} + \alpha_1 H_{N-1}}{|\boldsymbol{\alpha}|^2} \\
T_x = &   \frac{T_+ + T_-}{2}       \\  
T_y = &  \frac{T_+ - T_-}{2 \:i}      \\
T_\perp = & \frac{-\alpha_1 H_{N-2} + \alpha_2 H_{N-1}}{|\boldsymbol{\alpha}|^2}   \: .  \label{eqa_4}
\end{split}
\ee
Thus, $H_{N-1}$ and $H_{N-2}$ can be expressed as
\be
\begin{split}
  H_{N-1} = & \:\alpha_1 T_z + \alpha_2 T_\perp  \:,\\ 
  H_{N-2} = & \:\alpha_2 T_z - \alpha_1 T_\perp  \: .  
\label{eqa_5}
\end{split}
\ee
Make the ansatz that the Higgs field $\boldsymbol{\Phi}$ and vector potential $\boldsymbol{A}$   in Eq.~\ref{eq2_1} take the form
\be
\begin{split}
\boldsymbol{\Phi} = &  \:(  Q(r)  \: \alpha_1 T_z + \alpha_2 T_\perp ) \:v \:,   \\
\boldsymbol{A} = & \frac{g_e}{g} \:S(r) \:v \:\alpha_1   \:T_z    \:dt \:+ \:T_z \:(-C) \:W(r) (1-\cos\theta) \:d\phi \: , 
\end{split}
\label{eqa_6}
\ee
where 
\be
\begin{array}{lcr}
 W(r), Q(r), S(r) \:\rightarrow  0\:,       & \:\text{as} &  \:r \:\rightarrow 0 \:; \\
 W(r), Q(r) \:\rightarrow  1 \:,   \:S(r) \:\rightarrow  \:1 -  \frac{g} {g_m \:e \:\alpha_1 \:v \:r} \:,     & \:\text{as} & \:r \:\rightarrow \infty \:; \vspace{.0625in}  \\ 
\:g=   \sqrt{g_e^2 + g_m^2} \:. \label{eqa_22}
\end{array}
\ee
Here C is an arbitrary constant, and quantities $g_e$ and $g_m$ are the electric and magnetic charges.
Applying the gauge transformation 
\be
\chi  = e^{-i \phi T_z} e^{-i \theta T_y} e^{i \phi T_z} 
\label{eqa_15}
\ee 
to   $\boldsymbol{A}$ and   $\boldsymbol{\Phi}$  we obtain 
\be
\begin{split}
\boldsymbol{A}  \:\rightarrow & \:\chi \boldsymbol{A} \chi^{-1} - \frac{1}{i e}d\chi \:\chi^{-1} \\
   = &  \frac{g_e}{g}  \:S(r) \:v \:\alpha_1  \:T_r \:dt +  \:\frac{W(r)}{e} \:( \: T_\theta \: \sin \theta \:d\phi   \:- T_\phi \:d\theta \:) \:.
\end{split}
 \label{eqa_16}
\ee
and
\be
\begin{split}
\boldsymbol{\Phi} \:\rightarrow & \:\chi \boldsymbol{\Phi} \chi^{-1} \\
                    = & \:v    \:[\:\alpha_2 \:T_\perp +  \:Q(r)  \:\alpha_1 \:T_r ]  \:.
\end{split}
  \label{eqa_17}
\ee
We have used the fact that 
\be
d\chi \:\chi^{-1} = -i \:[(1- \cos\theta) \:T_r \:d\phi + \sin\theta \:T_\theta \:d\phi - \:T_\phi \:d\theta]  \:. \label{eqa_17a}  
\ee
The elements of the Lie algebra $T_r $, $T_\theta $, and $T_\phi$ are defined as (See Appendix~\ref{sect7}.)
\be
\begin{split}
T_r  = & \:T_x \:\sin \theta \cos \phi + T_y  \:\sin \theta \sin \phi 
              + T_z \:\cos \theta  \\
T_\theta = & \:T_x \:\cos \theta \cos \phi + T_y \:\:\cos \theta \sin \phi - T_z \:\sin \theta \\
 T_\phi  = & \:-T_x \:\sin \phi + T_y \:\cos \phi  \:.       
\end{split}
\label{eqa_17b}
\ee
In Eq.~\ref{eqa_6} the constant $C$ has been set equal to $1/e$  to eliminate the string singularity.   
For specificity we assume that $V(\boldsymbol{\Phi}\cdot \boldsymbol{\Phi})$ is given by Eq.~\ref{eq2_10}.  Following 't Hooft~\cite{thooftg76} we substitute Eqs~\ref{eqa_16} and \ref{eqa_17} into  the Lagrangian density, Eq.~\ref{eq2_1}, to obtain the Lagrangian  
\be
\begin{split}
L = \int d^3r \:\mathcal{L} =  
& 4 \pi\: \frac{g}{g_m\: |\boldsymbol{ \alpha}|^2 \:e}\: v\: \alpha_1   \:\bigg(
\:\frac{1}{2}   \int_{0}^{\infty} dx \:[ \{ \:s^{\prime \:2} \:x^2  + 2  \:s^2 \:(w-1)^2 \}(\frac{g_e}{g})^2 \\
  &  - \{2 \:w^{\prime \:2} +  \:\frac{w^2}{x^2} \:(w-2)^2\}(\frac{g_m}{g})^2   \\
  &  -  \{ \:q^{\prime \:2} \:x^2   + 2  \:q^2 \:(w-1)^2 \} 
   -\frac{2}{8}  \:\beta  \:x^2 \:(q^2-1)^2 ] \bigg) \:.
\end{split}
\label{eqa_7}
\ee
In Eq.~\ref{eqa_7} the variable of integration $r$ has been transformed to the dimensionless variable $x$ where
\be
x = \frac{g_m \:e \:v \:\alpha_1}{g}  \:r  \:. \label{eqa_8}
\ee
and 
\be
\beta = \:\lambda \left (\frac{g }{g_m\: e\: |\boldsymbol{\alpha}| }\right)^2 \:. \label{eqa_7a}
\ee
The functions $w$, $q$, and $s$ are the transformed functions $W$, $Q$, and $S$, i.e.
\be
\begin{split}
w(x) = &  \:W(r) \:,  \\
q(x) = & \:Q(r) \:,    \\
s(x) = &  \:S(r) \:.
\end{split}
\label{eqa_9}
\ee
In obtaining Eq.~\ref{eqa_7} we have used the relationships,
\be
\begin{split}
 \boldsymbol{F}_{t r} = &  \:\frac{g_e}{g}  \:S^\prime(r) \:v \:\alpha_1  \:T_r             \\
\boldsymbol{F}_{t \theta} = &  [1-W(r)] \:\frac{g_e}{g}  \:S(r) \:v \:\alpha_1   \:T_\theta   \\
\boldsymbol{F}_{t \phi} = &  [1-W(r)] \:\frac{g_e}{g}  \:S(r) \:v \:\alpha_1 \:\sin \theta\ \:T_\phi  \\
\boldsymbol{F}_{r \theta} = & -\frac{W^\prime(r)}{e} \:T_\phi    \\
\boldsymbol{F}_{r \phi} = &   \frac{W^\prime(r)}{e} \sin \theta  \:T_\theta  \\
\boldsymbol{F}_{\phi \theta} = &  \frac{W(r) \: (2-W(r))}{e} \:\sin \theta \:T_r   \:,
\end{split}
\label{eqa_10}
\ee
and
\be
\begin{split}
D_r \boldsymbol{\Phi} = &  Q^\prime(r) \:v  \: \alpha_1 \:T_r    \:  \\
D_{\theta} \boldsymbol{\Phi} = & [1-W(r)] \:Q(r) \:v \: \:\alpha_1 \:T_\theta  \\
D_{\phi} \boldsymbol{\Phi} = & [1-W(r)] \:Q(r) \:v \: \:\alpha_1 \:\sin \theta\ \:T_\phi    \:.
\end{split}
\label{eqa_11}
\ee
We now apply the variational principle to Eq.~\ref{eqa_7} with respect the functions $s(x)$, $q(x)$, and $w(x)$ to 
obtain the Euler-Lagrange equations:
\be
\begin{split}
(\frac{g_e}{g})^2 \{(x^2 s^\prime)^\prime - 2 s (w-1)^2 \} = 0  \\
(x^2 q^\prime)^\prime - 2 q (w-1)^2 - \frac{\beta}{2}x^2 (q^2-1) q = 0 \\
(\frac{g_m}{g})^2 \{ w^{\prime \prime} - \frac{w (w-1) (w-2)}{x^2} \} - (q^2 - (\frac{g_e}{g})^2 s^2 )(w-1) = 0 \: .
\end{split}
\label{eqa_12}
\ee

The magnetic and electric charges of the dyon can be obtained as follows.  The magnetic charge, $g_m$, is given by
\be
g_m = \int_{S_\infty} B^i dS_i \:, \label{eqa_13}
\ee
where\footnote{Fields given in standard font  correspond to electromagnetic fields while those in boldface correspond to Yang Mills fields, e.g.\@ $\boldsymbol{B}^i = -\frac{1}{2}\epsilon^{ijk} \boldsymbol{F}_{j k} $ or  $\boldsymbol{E}^i = - \boldsymbol{F}^{0 i}$. }
\be 
B^i = -\frac{1}{2}\epsilon^{ijk} F_{jk} = \frac{1}{e} \frac{\delta^{i r}}{r^2} \:.  \label{eqa_13a}
\ee
The quantity $F_{ij}$ is obtained from Eqs.~\ref{eq2_3} and \ref{eqa_10}.  Thus, asymptotically in the limit of  large $r$  
\be
\begin{split}
g_m & =  \int_{S_\infty} (\boldsymbol{B} \cdot \bar{\boldsymbol{\Phi}})^i dS_i  \\
   & = \frac{1}{v \alpha_1}\int_{S_\infty} (\boldsymbol{B} \cdot \boldsymbol{\Phi})^i dS_i \\
     & = \int_{S_\infty} B^i dS_i  \\ 
   & = \frac{4 \pi}{|\boldsymbol{\alpha}|^2  e}\:. 
\end{split}
\label{eqa_18}
\ee
We have used the fact that 
\be
T_a \cdot T_b = \frac{1}{|\boldsymbol{\alpha}|^2} \delta_{a b}  \:, \label{eqa_19}
\ee
for ($a,b = r, \theta, \phi$).
Similarly, the electric charge is given by
\be
\begin{split}
g_e = \frac{1}{4 \pi}\int_{S_\infty} E^i dS_i \:, \label{eqa_20}
\end{split}
\ee
where 
\be
E^i = -F^{0i}= \:\frac{g_e}{g |\boldsymbol{\alpha}|^2}  \:S^\prime(r) \:v \:\alpha_1  \delta^{ir}  \:. \label{eqa_20a}
\ee
 Thus, asymptotically in the limit of  large $r$   
\be
    E^i =    g_e \frac{\delta^{i r}}{r^2}                     \label{eqa_20b}
\ee 
so that
\be
\begin{split}
g_e & = \frac{1}{4 \pi} \int_{S_\infty} (\boldsymbol{E} \cdot \bar{\boldsymbol{\Phi}})^i dS_i  \\
   & = \frac{1}{4 \pi v \alpha_1}\int_{S_\infty} (\boldsymbol{E} \cdot \boldsymbol{\Phi})^i dS_i \\
     & = \frac{1}{4 \pi}\int_{S_\infty} E^i dS_i   \\ 
   & = g_e \:. 
\end{split}
\label{eqa_21}
\ee
The asymptotic form of $S(r)$ given by Eq.~\ref{eqa_22} was chosen specifically to yield this result.  Furthermore, the electric charge is quantized in integer multiples of the eigenvalues, $h_{N-1}$, of the operator $H_{N-1}$
\be
g_e = n h_{N-1} e \:,   \label{eqa_24}
\ee
where $n$ is an integer. For the fundamental representation $h_{N-1} = \frac{\alpha_1}{N}$ so that 
\be
g_e = n \frac{\alpha_1}{N} e\:.  \label{eqa_23}
\ee
Substituting for $\alpha_1$ (Eq.~\ref{eqa_2}) we obtain
\be
g_e= n\: \frac{1}{N}\: \sqrt{\frac{N}{2(N-1)}} \:  e  \:. \label{eqa_25}
\ee

We now derive an explicit expression for the mass of the dyon.  To facilitate this derivation we, first, express the magnetic charge, $g_m$, and electric charge, $g_e$ alternatively as 
\be
g_m =  \frac{1}{v \alpha_1}\int \boldsymbol{B}^i \cdot D^i \boldsymbol{\Phi} \:d^3r \:, \label{eqa_26} 
\ee
and
\be
g_e = \frac{1}{4 \pi}  \frac{1}{v \alpha_1}\int \boldsymbol{E}^i \cdot  D^i \boldsymbol{\Phi} \:d^3r \:, \label{eqa_27} 
\ee
Equations~\ref{eqa_26} and \ref{eqa_27} have been obtained by integrating Eqs.~\ref{eqa_18} and \ref{eqa_21} by parts and using the fact that $\boldsymbol{F}_{\mu \nu}$ satisfies the Bianchi identity and  Euler equation, i.e. (See, for example, Harvey~\cite{harveyj96} who discusses this in some detail.) 
\be
\begin{split}
 D_{[\alpha} \boldsymbol{F}_{\mu \nu ]} & = 0 \\
	 D_{[\alpha} \boldsymbol{^*F}_{\mu \nu]} & = 0  \:. \\
\end{split}
\label{eqa_28}
\ee

We now proceed with calculating the mass of the dyon as follows.   Since the field $\Phi$ does not depend on time, the energy (mass), $m_d$, of the system is given by 
\be
m_d = \frac{1}{2} \int [ \boldsymbol{E}^i \cdot \boldsymbol{E}^i +  \boldsymbol{B}^i \cdot\boldsymbol{B}^i +  D^i \boldsymbol{\Phi} \cdot D^i \boldsymbol{\Phi}  +    \frac{\lambda}{8} (\boldsymbol{\Phi} \cdot \boldsymbol{\Phi} -v^2)^2   ] \:d^3r \:. \label{eqa_29}
\ee
The mass can be expressed more conveniently as follows.  Define 
\be
\Delta^2(r) = (\boldsymbol{E}^i - \frac{g_e}{g} D^i \boldsymbol{\Phi}) \cdot (\boldsymbol{E}^i - \frac{g_e}{g} D^i \boldsymbol{\Phi})  + (\boldsymbol{B}^i - \frac{g_m}{g} D^i \boldsymbol{\Phi}) \cdot (\boldsymbol{B}^i - \frac{g_m}{g} D^i \boldsymbol{\Phi})  \:. \label{eqa_30}
\ee
Expanding Eq.~\ref{eqa_30}, substituting it into Eq.~\ref{eqa_29}, and performing the change of  variables, Eq.~\ref{eqa_8}, we obtain
\be
\begin{split}
m_d  & = 4 \pi\: \frac{g}{g_m\: |\boldsymbol{ \alpha}|^2 \:e}\: v\: \alpha_1   \:(1+
\:\delta^2_m ) \\
     & = g\: v\: \alpha_1   \:(1+\:\delta^2_m )  \:, 
\end{split}
\label{eqa_31}
\ee
where 
\be
\delta^2_m = \:\frac{1}{2} \int_{0}^{\infty} dx \:x^2 \:[ \delta^2(x)   + \frac{2}{8}  \:\beta   \:(q^2-1)^2 ] \: \label{eqa_31a}
\ee
and
$\delta^2(x) = \Delta^2(r)$.  Specifically,
\be
\begin{split}
\delta^2(x)  = & (\frac{g_m}{g})^2 \:[2\frac{(w^\prime-q(w-1))^2}{x^2} + (\frac{w(w-2)}{x^2} -q^\prime)^2] \\
  & +(\frac{g_e}{g})^2 [(s^\prime -q^\prime)^2 + 2 \frac{(w-1)^2 (s-q)^2}{x^2}] \:.
\end{split}
\label{eqa_32}
\ee
In obtaining Eq.~\ref{eqa_31} we have also used Eqs.~\ref{eqa_26} and \ref{eqa_27}.

In general, the Euler-Lagrange equations cannot be solved in closed form; however, if $V(\boldsymbol{\Phi} \cdot \boldsymbol{\Phi}) = 0$, i.e.\@ $\lambda = 0$,  one can show that $\boldsymbol{E}^i = \pm \frac{g_e}{g} D^i \boldsymbol{\Phi} $ and $\boldsymbol{B}^i = \pm \frac{g_m}{g} D^i \boldsymbol{\Phi} $ are exact solutions, or equivalently
\be
\begin{split}
s & =\pm  q  \\
w^\prime & = \pm q (w-1)  \\
x q^\prime & = \pm \frac{w (w-2) }{x} \:.
\end{split}
\label{eqa_33}
\ee
Equations~\ref{eqa_33} can be solved (See Harvey~\cite{harveyj96}.)
\be
\begin{split}
w & = 1 - \frac{x}{\sinh(x)}  \\
q & = \coth(x) - \frac{1}{x}  \:.
\end{split}
\label{eqa_34}
\ee
For this solution $\delta^2_m = 0$ so that the mass of the dyon assumes its minimum value
\be
m_d  = g v \alpha_1 \:. \label{eqa_35} 
\ee
This is a BPS state.

\begin{figure}
\begin{center}
\epsfig{file=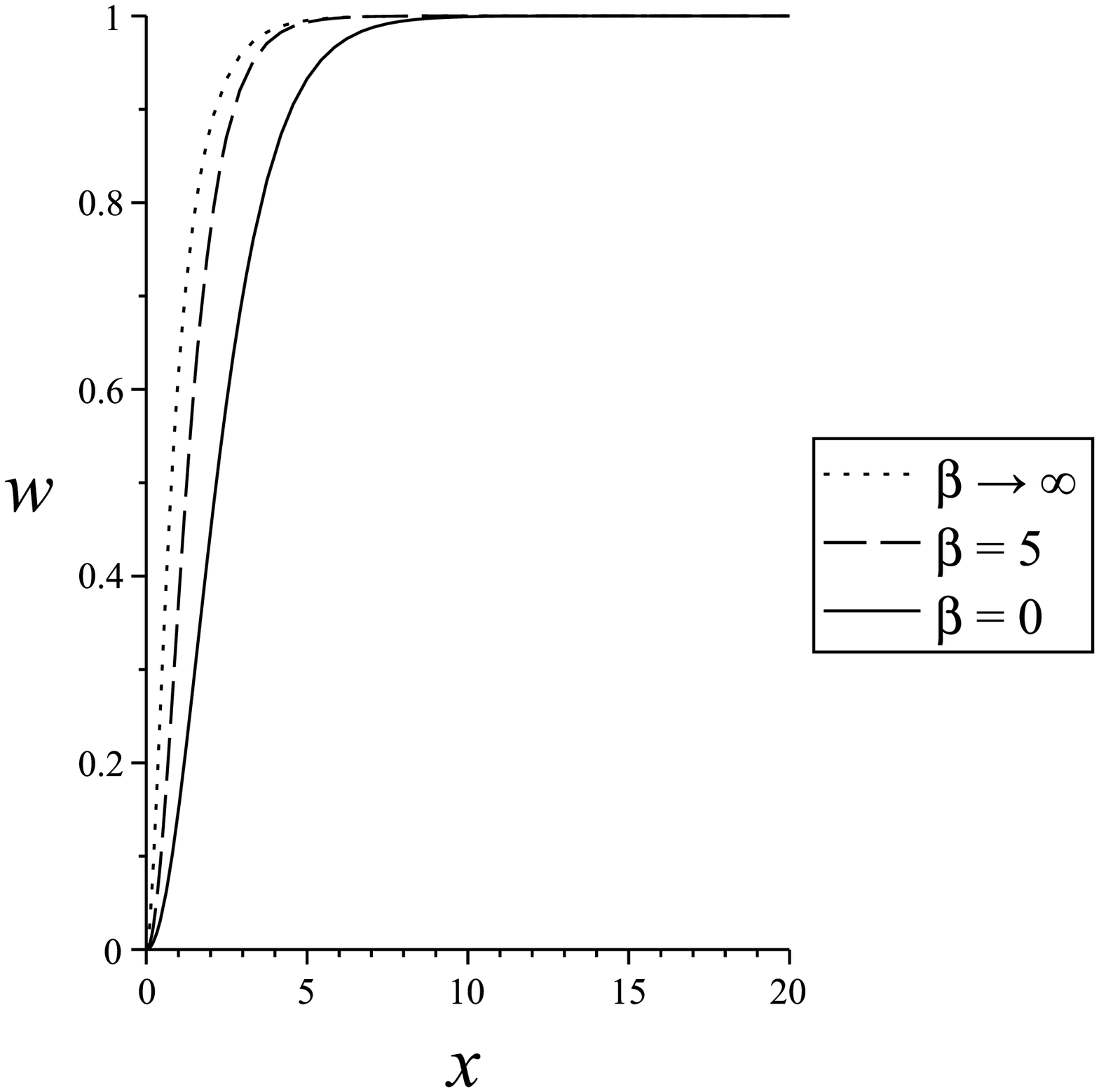,height=6cm} 
\end{center}
\caption{Shown are numerical solutions of the function $w(x)$ when $\beta =5$ and $\beta \rightarrow \infty$. For comparison the exact solution when $\beta = 0$ is also shown. }
\label{fig3}
\end{figure}

\begin{figure}
\begin{center}
\epsfig{file=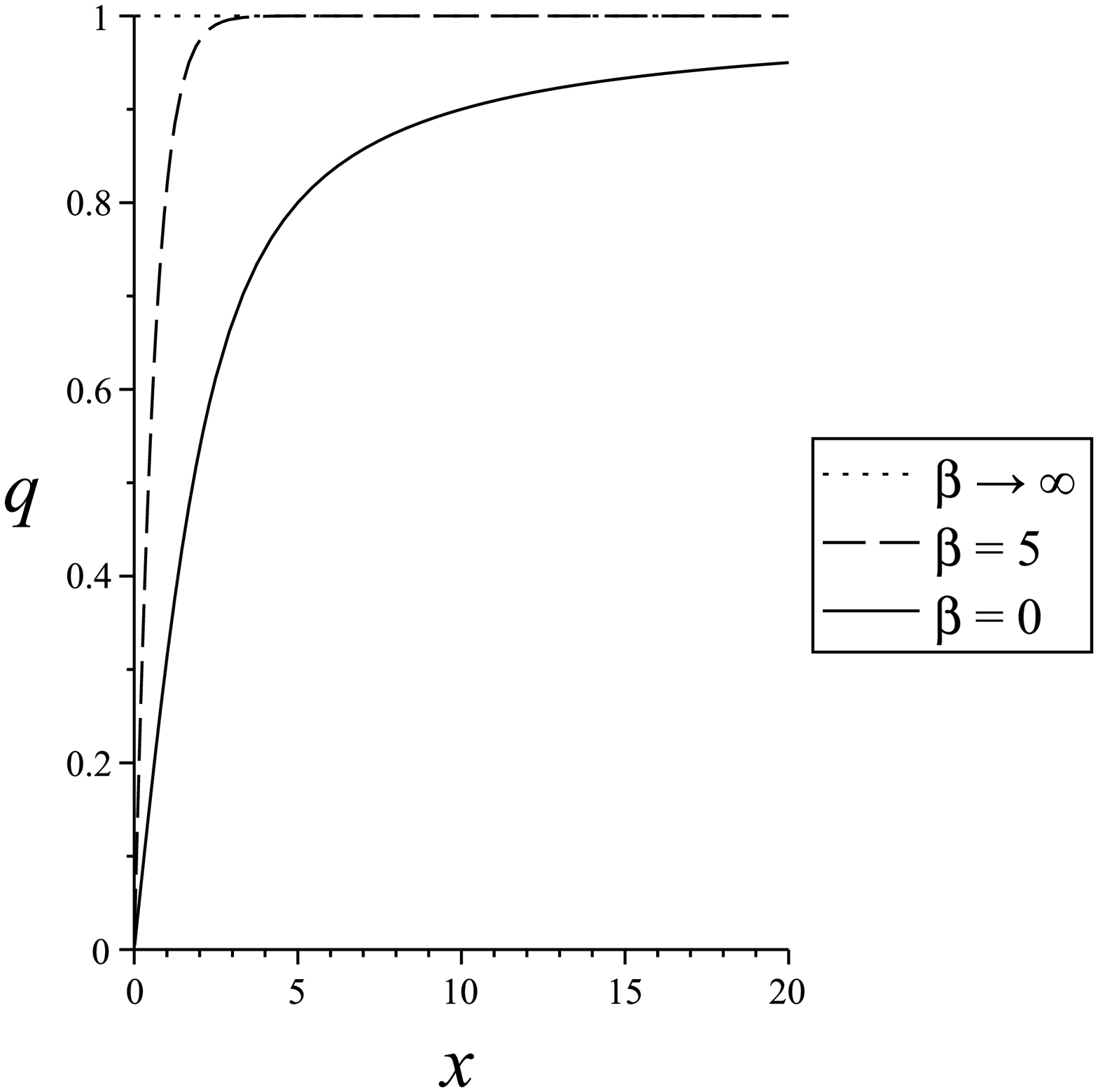,height=6cm} 
\end{center}
\caption{Shown are numerical solutions of the function $q(x)$ when $\beta =5$ and $\beta \rightarrow \infty$. For comparison the exact solution when $\beta = 0$ is also shown. }
\label{fig4}
\end{figure}

\begin{figure}
\begin{center}
\epsfig{file=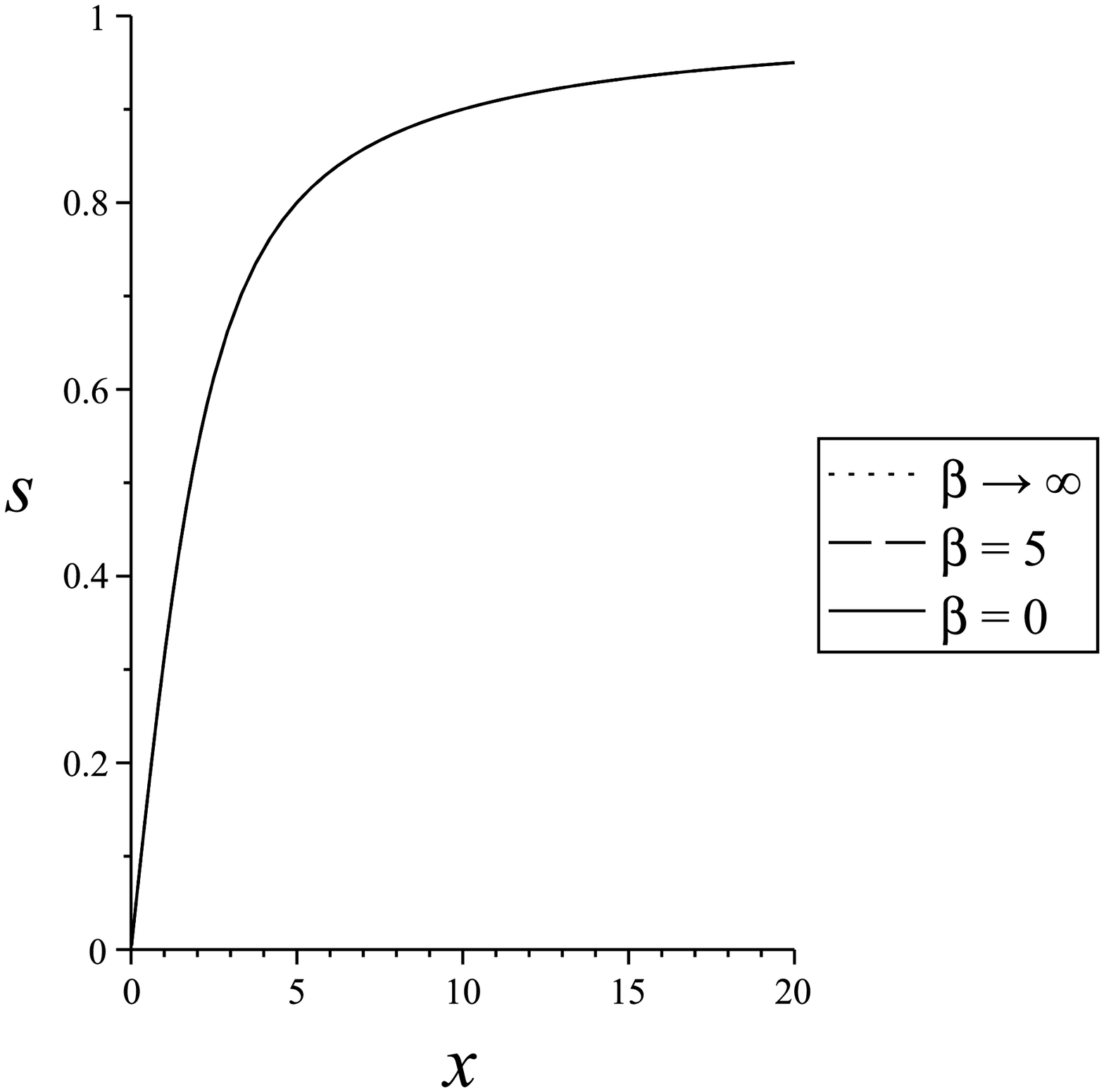,height=6cm} 
\end{center}
\caption{Shown are numerical solutions of the function $s(x)$ when $\beta =5$ and $\beta \rightarrow \infty$. For comparison the exact solution when $\beta = 0$ is also shown. To the accuracy of the approximations the numerical solutions are indistinguishable from the exact solution corresponding to $\beta=0$.}
\label{fig5}
\end{figure}

General solutions to Eqs.~\ref{eqa_12} exhibit the following behavior:
\be
\begin{array}{lcr}
 w(x) \:\sim x^2, \:q(x)\:\sim  x,   \:\text{and}  \:s(x) \:\sim  x,       & \:\text{as} &  \:x \:\rightarrow 0 \:; \\
 w(x) \:\sim 1, \:q(x) \:\sim 1-\frac{\exp(-\sqrt{\beta}x)}{x}, \:\text{and}    \:s(x) \:\sim  \:1 -  \frac{1}{x} \:,     & \:\text{as} & \:x \:\rightarrow \infty \:. 
\end{array}
\label{eqa_36}
\ee
In Figures~\ref{fig3}, \ref{fig4}, and \ref{fig5} we show numerical solutions for $(\frac{g_e}{g})^2 \approx (\frac{g_m}{g})^2 \approx \frac{1}{2}$ when  $\beta =5$ and $\beta \rightarrow \infty$. For comparison we also show  the solution $\beta=0$, i.e.\@ $\lambda=0$.  
These two numerical solutions can be used for estimating the quantity $\delta^2_m$ on whose value the the mass of the dyon depends. Of relevance in performing these calculations and not apparent from Figure~\ref{fig4}  is that as  $\beta \rightarrow \infty$,     $q \approx \sqrt{\beta} x$  for $ x << 1/\sqrt{\beta}$.   Consequently, we can integrate  Equation~\ref{eqa_31a}, numerically, to obtain
\be
\delta^2_m  \approx 
\begin{cases} 
  .41 & \text{for}  \;\beta=5,  \\
 .63 & \text{as}  \;\beta \rightarrow \infty  \:.
\end{cases}
\label{eqa_37}
\ee
These results suggest that the mass of the dyon is relatively insensitive to the value of $\beta$, a result alluded to in the work of 't Hooft.\cite{thooftg76}

\setcounter{equation}{0}
\section{Application to SU(3) and G2}
\subsection{SU($3$)}
\label{sect2_2}

\begin{figure}
\begin{center}
\begin{pspicture}(-4.2,-3.9)(4.2,3.9)
%\psgrid[subgriddiv=1,griddots=10]
\qline(-4.2,0)(4.2,0)
\uput[0](4.15,-.25){$H_1$}
\qline(0,-3.9)(0,3.9)
\uput[0](.1,4.0){$H_2$}
\SpecialCoor
\psline[linestyle=dashed,linecolor=black](0,-.05)(-4.2,-.05)
\uput[0](-4.95,-.25){$H^{\prime \prime}_1$}
\psline[linestyle=dashed, linecolor=black](.05,0)(.05,-3.9)
\uput[0](.2,-3.9){$H^{\prime \prime}_2$}
\psline[linestyle=dotted,linecolor=black](.2,0)(4.2;120)
\uput[0](4.2;119){$H^{\prime}_1$}
\psline[linestyle=dotted,linecolor=black](0,0)(4.2;210)
\uput[0](4.2;201){$H^{\prime}_2$}
\pscircle[fillcolor=black,fillstyle=solid,linecolor=black](-0.1,0.0){.1}
\pscircle[fillcolor=black,fillstyle=solid,linecolor=black](+0.1,0.0){.1}
\pscircle[fillcolor=black,fillstyle=solid,linecolor=black](+2.0,3.46){.1}
\uput[0](1.7,2.9){$(\frac{1}{2},\frac{\sqrt{3}}{2})$}
\pscircle[fillcolor=black,fillstyle=solid,linecolor=black](-2.0,3.46){.1}
\uput[0](-3.4,2.9){$(-\frac{1}{2},\frac{\sqrt{3}}{2})$}
\pscircle[fillcolor=black,fillstyle=solid,linecolor=black](+2.0,-3.46){.1}
\uput[0](1.7,-2.9){$(\frac{1}{2},-\frac{\sqrt{3}}{2}) $}
\pscircle[fillcolor=black,fillstyle=solid,linecolor=black](-2.0,-3.46){.1}
\uput[0](-3.4,-2.9){$ (-\frac{1}{2}, -\frac{\sqrt{3}}{2})$}
\pscircle[fillcolor=black,fillstyle=solid,linecolor=black](+4.0,0){.1}
\uput[0](3.0,.3){$(1, 0)$}
\pscircle[fillcolor=black,fillstyle=solid,linecolor=black](-4.0,0){.1}
\uput[0](-4.2,.3){$(-1, 0)$}
\end{pspicture}
\end{center}
\caption{Root system of SU(3). The axes $H_1$ and $H_2$ show the angles that the root vectors make with respect to the two elements of the Cartan subalgebra. The primed and double primed axes represent two other equivalent sets of axes, one corresponding to the center of the group SU(3) and the other to the group SU(2). }
\label{fig1}
\end{figure}
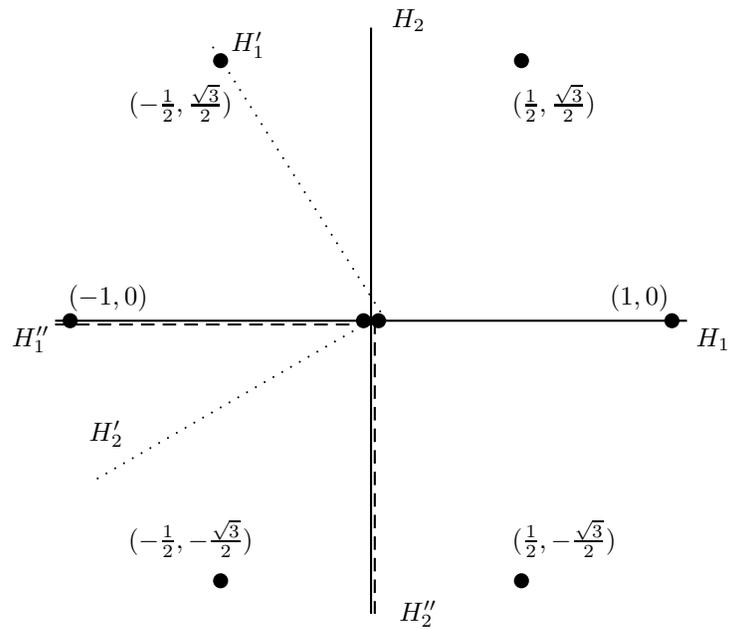

As a concrete application of the theory presented in Section~\ref{sect5} we apply the theory to SU(3) and contrast  the dyon solutions obtained to those of SU(2).  Our purpose is to construct those dyon solutions which are inherent to SU(3) while not being associated with those of the  various SO(3) or SU(2) subgroups of SU(3). 
The generators of SU(3) are $T_a$, ($T_a= \lambda_a/2, \lambda_a$ being the Gell-Mann matrices).  The Cartan subalgebra $H_{3-i}$ ($i=2, 1$) is $H_1 = T_3$ and $H_2 = T_8$. First, we assume that the Higgs field, $\boldsymbol{\Phi}$, asymptotically approaches  the vacuum state of the potential given by Eq.~\ref{eq2_10}  for  large values of the radial coordinate $r$, i.e.\@ $\lim_{r\rightarrow\infty} \boldsymbol{\Phi} \cdot \boldsymbol{\Phi} = v^2$.  Figure~\ref{fig1} is a depiction of the root system of SU(3). Based on this figure an obvious choice for the asymptotic form of $\boldsymbol{\Phi}$ is
\be
\boldsymbol{\Phi} = v H_2 \:, \label{eq2_9a} 
\ee
so that $\boldsymbol{\bar{\Phi}}$ is 
\be
\boldsymbol{\bar{\Phi}}  = \frac{1}{\alpha_1 v} \boldsymbol{\Phi} = \frac{H_2}{\alpha_1} \: . \label{eq2_9b}
\ee
where $\alpha_1 = \frac{\sqrt{3}}{2}$ for SU(3).  This choice of $\boldsymbol{\bar{\Phi}}$ can be seen to satisfy Equation~\ref{eq2_8} by inspection of Figure~\ref{fig1}.  There are other equivalent choices for  $\boldsymbol{\bar{\Phi}}$, which can be obtained by $2\pi/6$ rotations of the coordinate axes of the root diagram,  $H^\prime_2$ and $H^{\prime \prime}_2$ being two such examples.  They are related to the center of SU(3) and SU(2), respectively~\cite{sallerh98}.  Thus, there are $3 \times 2$  equivalent choices for $\boldsymbol{\bar{\Phi}}$. An equivalent way of understanding the factor of three is to note that $H_2$ is a diagonal matrix with values of one in all diagonal elements except for a value of minus two in one of the diagonal elements; however, the minus two can be in any one of the three diagonal elements resulting in three possibilities of $H_2$. This result generalizes to SU($N$) in the obvious way.

For SU(3)  the simple root vector $E_\alpha$ corresponding to   Eq.~\ref{eqa_1} is $\boldsymbol{\alpha}=(\alpha_2, \alpha_1)=(\frac{1}{2}, \frac{\sqrt{3}}{2})$. The gauge transformation, $\chi$, used to remove the string singularity is given by (For comparison see, for example, the discussion of Ryder~\cite{ryderl96}.) 
\be
\begin{split}
\chi & = e^{-i \phi T_z} e^{-i \theta T_y} e^{i \phi T_z} \\
  & = \begin{pmatrix}
  \cos \theta/2 & 0 & - e^{-i\phi}  \sin \theta /2 \\
   0 & 1 & 0 \\
e^{i\phi}   \sin \theta/2 & 0 & \cos \theta/2 
\end{pmatrix}  \: . 
\end{split}
\label{eq2_15}
\ee 
Applying the results of Section~\ref{sect5} and using the fact that $|\boldsymbol{\alpha}|=1$ we obtain the following results.  The magnetic field of the dyon is 
\be
B^i = \frac{1}{e} \frac{\delta^{i r} }{r^2} \:.  \label{eq2_16}
\ee
Asymptotically, as $r \rightarrow \infty$, the electric field is
\be
E^i = g_e \frac{\delta^{i r}}{r^2}  \:,  \label{eq2_17}
\ee 
where, in the case of the fundamental representation,
\be
g_e = n\: \frac{1}{3}\frac{\sqrt{3}}{2}e   \:. \label{eq2_18}
\ee
The mass of the dyon is given by
\be
  m_d =   g\: v\: \frac{\sqrt{3}}{2}   \:(1+\:\delta^2_m )        \:. \label{eq2_19}
\ee
The SU(3) dyon is, in a certain sense,  less massive than the corresponding SU(2) dyon by a factor of $\frac{\sqrt{3}}{2}$, since  $\alpha_1 = 1$ for SU(2). The electric charge and magnetic charge satisfy the relationship
\be
g_e \: g_m =  4 \pi \frac{n}{3}\frac{\sqrt{3}}{2} \:.  \label{eq2_20}
\ee
For comparison with SU(2) the relationship is
\be
g_e \: g_m =  4 \pi \frac{n}{2} \:.  \label{eq2_21}
\ee

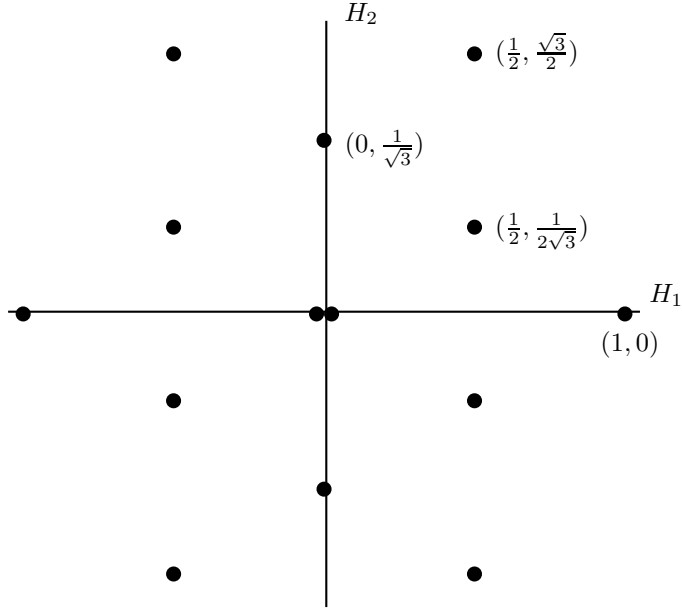
\begin{figure}
\begin{center}
\begin{pspicture}(-4.2,-3.9)(4.2,3.9)
%\psgrid[subgriddiv=1,griddots=10]
\qline(-4.2,.03)(4.2,.03)
\uput[0](4.15,.25){$H_1$}
\qline(.03,-3.9)(.03,3.9)
\uput[0](.1,4.0){$H_2$}
\SpecialCoor
\pscircle[fillcolor=black,fillstyle=solid,linecolor=black](-0.1,0.0){.1}
\pscircle[fillcolor=black,fillstyle=solid,linecolor=black](+0.1,0.0){.1}
\pscircle[fillcolor=black,fillstyle=solid,linecolor=black](+2.0,3.46){.1}
\uput[0](2.10,3.5){$(\frac{1}{2}, \frac{\sqrt{3}}{2})$}
\pscircle[fillcolor=black,fillstyle=solid,linecolor=black](-2.0,3.46){.1}
\pscircle[fillcolor=black,fillstyle=solid,linecolor=black](+2.0,-3.46){.1}
\pscircle[fillcolor=black,fillstyle=solid,linecolor=black](-2.0,-3.46){.1}
\pscircle[fillcolor=black,fillstyle=solid,linecolor=black](+4.0,0){.1}
\uput[0](3.5,-.4){$(1, 0)$}
\pscircle[fillcolor=black,fillstyle=solid,linecolor=black](-4.0,0){.1}
\uput[0](.1,2.2){$(0, \frac{1}{\sqrt{3}})$}
\pscircle[fillcolor=black,fillstyle=solid,linecolor=black](2.31;30.0){.1}
\uput[0](2.10,1.1){$(\frac{1}{2}, \frac{1}{2 \sqrt{3}})$}
\pscircle[fillcolor=black,fillstyle=solid,linecolor=black](+2.31;90.0){.1}
\pscircle[fillcolor=black,fillstyle=solid,linecolor=black](+2.31;150.0){.1}
\pscircle[fillcolor=black,fillstyle=solid,linecolor=black](2.31;210.0){.1}
\pscircle[fillcolor=black,fillstyle=solid,linecolor=black](+2.331;270.0){.1}
\pscircle[fillcolor=black,fillstyle=solid,linecolor=black](2.31;330.0){.1}
\end{pspicture}
\end{center}
\caption{Root system of G2. The axes $H_1$ and $H_2$ show the angles that the root vectors make with respect to the two elements of the Cartan subalgebra.}
\label{fig2}
\end{figure}

At this point we comment that, substantively, there is little difference between SU(3) and SU(2) monopoles, other than the difference in their mass.  Another difference relates to the interpretation of   Eq.~\ref{eqa_6}. Asymptotically, in  the SU(2) case $\boldsymbol{\Phi}$ is a mapping from the two sphere in configuration space into a two sphere of radius $v$ in field space; whereas, in the SU(3) case  $\boldsymbol{\Phi}$ is a mapping of the two-sphere  in configuration into a two-sphere of radius $v \: \frac{\sqrt{3}}{2}$ in field space.

\subsection{G2}
\label{sect2_2b}
For G2 it is not possible to apply the definition of the electromagnetic field, Eq.~\ref{eq2_3}, to an arbitrary gauge field for a particular field $\boldsymbol{\bar{\Phi}}$, or equivalently  there does not exist a field $\boldsymbol{\bar{\Phi}}$  satisfying the condition Eq.~\ref{eq2_8}, as is apparent from 
 studying the root system of G2 depicted in Figure~\ref{fig5}. If, however, we restrict our consideration to two families of gauge fields that are linear combintations of the Cartan subalgebra and either the long root vectors or the short root vectors,  it is then  possible to find  two fields $\boldsymbol{\bar{\Phi}}$ which  satisfy the condition Eq.~\ref{eq2_8}, one for each family of  gauge fields.  These two fields are
\be
\boldsymbol{\bar{\Phi}_i} =
\begin{cases}
    \frac{H_1}{\alpha_2}  & \text{if $i=1$},  \\
    \frac{H_2}{\alpha_1}  & \text{if $i=2$}.  
\end{cases}
\label{eq2b_1}
\ee
Decomposition of the adjoint representation in this manner is related to the fact that SU(3) is a regular and maximal subalgebra of G2~\cite{georgih82}.  Specifically, the long root vectors and the Cartan subalgebra of G2 form an SU(3) subalgebra of G2.  Under this SU(3) algebra the 14 dimensional adjoint representation of G2 transforms as an $8 \oplus 3 \oplus\bar{3}$. In addition, there is an SU(2) subalgebra of G2 which tranforms each element $E_{\boldsymbol{\alpha}}$ of the $3$ representation   into an element $E_{-\boldsymbol{\alpha}}$ of the $\bar{3}$.  Thus, we have two different electric fields, one associated with the long root vectors and $\boldsymbol{\bar{\Phi}_2}$  and the other associated with the short root vectors and $\boldsymbol{\bar{\Phi}_1}$.  The dyon solutions associated with $\boldsymbol{\bar{\Phi}_2}$ are the SU(3) dyon solutions discussed in Section~\ref{sect2_2}, while the dyon solutions associated with $\boldsymbol{\bar{\Phi}_1}$ possess different properties (We refer to these dyons as e-dyons to distinguish them from the dyons associated with the long root vectors).  Although The e-dyon solutions  cannot be obtained directly from the results presented in Section~\ref{sect5}, they can be obtained from those results with minor modification, i.e.\@ interchange 1 with 2 in Eqs.~\ref{eqa_1} and \ref{eqa_2}, and then proceed with the analysis, as we now describe briefly.  Consider the short root vector $\boldsymbol{\alpha} = (\frac{1}{2},\frac{1}{2 \:\sqrt{3}})$, and define 
\be
 \boldsymbol{\bar{\Phi}} = \boldsymbol{\bar{\Phi}_1}  = \frac{H_1}{1/2}                       \:.  \label{eq2b_2}
\ee
Proceeding with the analysis as in Section~\ref{sect2_2}, we obtain the following results. Since $|\boldsymbol{\alpha}|^2 =1/3$,  the e-dyon's magnetic charge is
\be
g_m = 4 \pi \frac{3}{e}    \:. \label{eq2b_3}
\ee
In addition, the quantization condition for the electric charge is 
\be
g_e = \frac{n}{2}   \:. \label{eq2b_4}
\ee
Thus, the relationship satisfied by the magnetic and electric charges is
\be
 g_e g_m = 4\pi \:\frac{n}{2} \:3 \:. \label{eq2b_5}
\ee
Finally, the mass of the e-dyon is
\be
m_d = \frac{g v}{2}  \:(1+\:\delta^2_m )   \:.   \label{eq2b_6}        
\ee
\setcounter{equation}{0}
\section{Conclusions}
\label{sect6}
In Section~\ref{sect2_1} we  have adopted the definition  of the electromagnetic field  first proposed by  't~ Hooft~\cite{thooftg76} in the context of SO(3). In the application of this definition to other gauge groups  we   have suggested that a reasonable criterion, which should be satisfied, is  that  the electromagnetic field defined in this manner should exist for an arbitrary gauge field.   We then have derived a specfic condition, Eq.~\ref{eq2_8}, which is necessary for this criterion to be satisfied. Applying the definition of the electromagnetic field to SU($N$) in Section~\ref{sect5} we have constructed dyon solutions possessing both topological electric  and magnetic charge.  Assuming a $|\boldsymbol{\Phi}|^4$--\:like potential for the Higgs field we  have estimated the mass of the dyon, finding it   to be relatively insensitive to the coupling parameter $\lambda$ characterizing the potential and only slightly greater than  the BPS bound.  Finally, we have applied  the general results of Section~\ref{sect5} specifically to SU(3) and G2.  For SU(3) the electric/magnetic charge relationship and mass of the dyon are given by Eqs.~\ref{eq2_20} and \ref{eq2_19}. For G2 we have found that it is not possible to satisfy the  criterion for the electromagnetic field; however, considering G2 under the action of its SU(3) subalgebra and relaxing the criterion imposed on the electromagnetic field  we  have discovered two different types of dyon solutions. One of these solutions corresponds  to dyon solutions associated with SU(3).  The other solution, denoted an e-dyon, has somewhat atypical properties.  Most notable is the fact that the magnetic charge   is $g_m = 4 \pi\: 3/e$,  $e$ being the gauge coupling.  This differs from the 't Hooft/Polyakov monople where $g_m = 4 \pi\: 1/e$.
\renewcommand{\theequation}{A-\arabic{equation}}
\setcounter{equation}{0}
\appendix
\section{Appendix}
\label{sect7}
Herein, we provide mathematical relationships which are useful in deriving results presented in Section~\ref{sect5}.  The quantities $T_i$, ($i = r, \theta, \phi$), are a representation of the SU(2) algebra, and $T_\perp$ commutes with each of the $T_i$, i.e.
\be
\begin{split}
T_i \wedge T_j & = i \epsilon_{ijk} T_k  \\
T_\perp \wedge T_i  & = 0   \:.
\end{split}
\label{eq_appendix_1}
\ee 
Furthermore,
\be
\begin{split}
\text{Tr($T_i T_j$)} & = \frac{1}{|\boldsymbol{\alpha}|^2} \frac{1}{2} \:\delta_{i j}  \\
\text{Tr($T_\perp T_\perp$)} & = \frac{1}{|\boldsymbol{\alpha}|^2} \frac{1}{2} \\
 \text{Tr($T_\perp T_i$)} & =    0                                    \:.
\end{split}
\label{eq_appendix_2}
\ee

\newpage
\bibliography{manuscript}  
\end{document}